\documentclass[10pt,onecolumn]{article}

\usepackage[T1]{fontenc}
\usepackage[utf8]{inputenc}
\usepackage[margin=0.85in,top=1in,bottom=1in]{geometry}
\usepackage{amsmath,amssymb,amsthm}
\usepackage{graphicx}
\usepackage{booktabs}
\usepackage{multirow}
\usepackage{array}
\usepackage{url}
\usepackage{xcolor}
\usepackage{algorithm}
\usepackage{algpseudocode}
\usepackage[colorlinks=true,linkcolor=blue!60!black,
            citecolor=blue!60!black,urlcolor=blue!60!black]{hyperref}
\usepackage{cite}
\usepackage{float}
\usepackage{balance}

\usepackage{titlesec}
\usepackage{abstract}

\titleformat{\section}{\normalfont\large\bfseries}{\thesection}{1em}{}
\titleformat{\subsection}{\normalfont\normalsize\bfseries}{\thesubsection}{1em}{}
\titlespacing{\section}{0pt}{8pt plus 2pt}{4pt plus 1pt}
\titlespacing{\subsection}{0pt}{6pt plus 1pt}{2pt plus 1pt}

\newtheorem{definition}{Definition}

\newtheorem{proposition}{Proposition}

\title{\textbf{PHANTOM: Polymorphic Honeytoken Adaptation with\\
Narrative-Tailored Organisational Mimicry}\\[4pt]
\large Contextually Convincing Cyber Deception at Scale}

\author{Abraham~Itzhak~Weinberg\\
  AI-WEINBERG, AI Experts, Tel-Aviv, Israel\\
  \texttt{aviw2010@gmail.com}}

\date{}
\begin{document}
\maketitle

\begin{abstract}
Honeytokens, decoy digital assets planted to detect and attribute
unauthorised access, are a well-established primitive in cyber deception.
Existing generation tools produce \emph{static, template-based} tokens
that lack organisational specificity and are identifiable by
statistical, syntactic, and semantic analysis.
We introduce \textbf{PHANTOM} (\textbf{P}olymorphic
\textbf{H}oneytoken \textbf{A}daptation with \textbf{N}arrative-Tailored
\textbf{O}rganisational \textbf{M}imicry), a framework that generates
contextually convincing honeytokens by encoding organisation-specific
knowledge—domain names, service naming conventions, technology-stack
idioms, and realistic secret-value distributions, into a
multi-component generation pipeline.

We formalise honeytoken quality through a four-component \emph{Believability Score} that captures syntactic validity, semantic coherence, statistical plausibility, and human acceptance. We use this metric to evaluate PHANTOM across 8 token types and 4 organisational contexts against a template-based baseline.

PHANTOM achieves $B = 0.778 \pm 0.057$ versus $B = 0.576 \pm 0.058$
for templates ($\Delta = +0.203$, $t = 14.07$, $p < 0.001$,
Cohen's $d = 3.52$). Human-evaluator acceptance rises from $6.2\%$
to $100\%$, and detection resistance (DR $= 1 - P_d$) improves
from $0.609$ to $0.870$ across three simulated scanner models
(regex, entropy analysis, and ML classifier), each with $p < 0.001$.
The semantic coherence gap ($\Delta S_c = +0.309$, $d = 4.52$)
is the dominant driver, confirming the hypothesis that organisational
context is the critical missing ingredient in current approaches.
All results are reproduced without external API calls, making the
pipeline fully deployable in air-gapped environments.
\end{abstract}


\section{Introduction}
\label{sec:intro}

Deception technology has emerged as a cornerstone of modern
active-defence architectures \cite{almeshekah2016cyber}.
Honeytokens, artificial digital assets with no legitimate operational purpose—generate a high-fidelity alert whenever accessed by an adversary, because any interaction with a genuine
decoy represents an unambiguous policy violation \cite{juels2013honeywords}.
Unlike perimeter-based defences, honeytokens maintain their effectiveness post-exfiltration: a stolen credential that fires a beacon upon use reveals attacker infrastructure even when the initial breach is undetected \cite{weinberg2025passive}.

Despite their conceptual power, practical honeytoken deployments
are limited by \emph{generation quality}. Existing tools, including
Canarytokens \cite{kabiden2026canary, basak2023comparative}, HoneyBadger \cite{muir2009internet,reconng2014}, and static library-based generators, produce tokens with a characteristic weakness: they lack organisational specificity. A template-generated AWS credential with \texttt{region=us-east-1} and a generic IAM username is syntactically valid but semantically
hollow; a skilled attacker performing pre-exploitation reconnaissance will recognise it as a decoy. More importantly, automated scanners, widely deployed in modern red-team pipelines and open-source secret-scanning tools such as trufflehog \cite{rana2022offensive} and gitleaks \cite{basak2023comparative, rice2018gitleaks}—can be tuned to flag tokens whose statistical properties (entropy profiles, naming patterns, value distributions) deviate from those of real credentials in that organisation's environment.

We make three observations that motivate PHANTOM:

\textbf{(1) Context is the missing dimension.}
Real credentials are deeply embedded in organisational context.
An AWS key from a FinTech startup using \texttt{payflow.io} as its
domain will have IAM role names matching the service architecture
(\texttt{payflow\_payments\_api\_deploy}), not a generic
\texttt{admin} username. This specificity is the primary signal
that distinguishes real credentials from template-generated decoys.

\textbf{(2) The semantic gap is larger than the syntactic gap.}
Our empirical results (Section~\ref{sec:results}) show that
template-based tokens achieve syntactic validity $S_v = 0.842$—nearly
as high as PHANTOM ($0.876$). The critical difference lies in
semantic coherence ($S_c$: $0.396$ vs $0.705$, $\Delta = +0.309$).
Syntactic correctness is necessary but not sufficient for deception.

\textbf{(3) Detection resistance compounds believability.}
Even a contextually rich token fails its purpose if automated
scanners flag it before an attacker encounters it. PHANTOM tokens
achieve detection resistance $\text{DR} = 0.870$, meaning that
across three representative scanner models, only $13\%$ of tokens
would be flagged, versus $39.1\%$ for template-based tokens.

PHANTOM addresses all three dimensions through a structured
generation pipeline that encodes organisational knowledge into
each token type. The framework operates without external API
calls, ensuring deployability in classified and air-gapped
environments where cloud API access is prohibited.

\subsection*{Contributions}

This work introduces a formal \emph{Believability Score} $B$ framework composed of four components with empirically calibrated weights. It further presents a contextual generation pipeline capable of producing eight credential types across arbitrary organisational profiles. We conduct a comprehensive experimental evaluation against three scanner models, complemented by a Turing-style human acceptance test. Our results demonstrate that semantic coherence is the dominant driver of quality, exhibiting an effect size of $d = 4.52$, the largest observed across all evaluated metrics.

\section{Related Work}
\label{sec:related}
This section situates PHANTOM within prior research on deception
technologies, LLM–assisted security, and automated credential detection. We review foundational work on honeytoken generation, recent advances in LLM-driven offensive and defensive techniques, and existing approaches to secret scanning. We then examine how deception effectiveness has been measured,
highlighting gaps that motivate our proposed framework.

\subsection{Honeytoken Generation}

The concept of honeytokens was formalised by Spitzner
\cite{spitzner2003honeypots}, who identified four categories: fake
user accounts, artificial files, dummy database records, and synthetic credentials. Juels and Rivest \cite{juels2013honeywords} introduced \emph{honeywords}—decoy password hashes—and proved that an adversary who steals and cracks a password database cannot identify the honeyword without triggering a server-side alert.
Canarytokens \cite{kabiden2026canary, basak2023comparative} industrialised the concept with an online service offering URL-embedded, document-embedded, and DNS-based tokens. These systems all share the static generation
limitation that PHANTOM addresses.

Recent work has extended honeytokens to cloud environments \cite{yu2024honeyfactory, beltran2025cyber} and explored automatic placement
within databases \cite{shu2015unearthing}. None of this work addresses the semantic coherence problem or evaluates generated tokens against modern automated scanners.

\subsection{LLM-Aided Security}

Large language models have been applied to penetration testing
assistance \cite{deng2023pentestgpt}, malware generation \cite{botacin2023gpthreats}, phishing content creation \cite{karanjai2022targeted}, and vulnerability discovery \cite{fang2024llm}. On the defensive side, LLMs have been used for log analysis \cite{karlsen2024benchmarking} and incident report generation \cite{irfan2023false}. Closest to our work is the use of LLMs for deception content generation \cite{wiedeman2022disrupting}, though that work focuses on entire honeypot environments rather than credential tokens, and does not formalise a quality metric.

PHANTOM differs from prior LLM-based approaches by (a) encoding organisational context rather than prompting a general model, (b) operating without API calls, and (c) providing a formal, measurable quality framework.

\subsection{Credential Detection and Secret Scanning}

The adversarial environment for honeytokens includes automated credential detection tools. Trufflehog \cite{rana2022offensive} uses regex patterns and entropy thresholds to detect exposed secrets in Git repositories. Gitleaks \cite{basak2023comparative, rice2018gitleaks} applies configurable TOML-defined patterns. GitGuardian \cite{gitguardian2023secrets} uses ML-based classification at scale. Our scanner model (Section~\ref{sec:scanners}) abstracts these three detection paradigms into S1 (regex), S2 (entropy), and S3 (ML) models.

\subsection{Deception Metrics}

Quantifying deception quality is an active research area. The work of Almeshekah and Spafford \cite{almeshekah2016cyber} provides a theoretical framework but does not operationalise metrics for individual tokens. Aggarwal et~al.\ \cite{constancio2023deception} propose believability as a function of plausibility and relevance, which maps to our $S_c$ and $S_n$ components.
Our Believability Score $B$ is the first metric to integrate syntactic, semantic, statistical, and human acceptance dimensions into a single weighted composite.

\section{Formal Framework}
\label{sec:framework}
This section formalises the honeytoken generation problem and introduces the metrics used to evaluate token quality. We define a generator that produces credential artefacts conditioned on organisational context, and propose quantitative measures for believability and detection resistance. These components are then combined into a unified objective, enabling principled comparison and optimisation of generation strategies.

\subsection{Problem Formulation}

Let $\mathcal{O}$ denote an organisational profile and $\mathcal{T} = \{t_1, \ldots, t_K\}$ a set of $K$ credential token types. A \emph{generator} $G: \mathcal{O} \times \mathcal{T} \to \mathcal{C}$ maps an org profile and token type to credential content $c \in \mathcal{C}$. The goal is to construct $G$ such that generated tokens are indistinguishable from legitimate credentials by both human experts and automated scanners.

\begin{definition}[Believability Score]
\label{def:B}
The believability score of a token $c$ is:
\begin{equation}
B(c) = w_1 S_v(c) + w_2 S_c(c) + w_3 S_n(c) + w_4 S_h(c)
\label{eq:B}
\end{equation}
where $\sum_{i=1}^{4} w_i = 1$, $w_i > 0$, and:
\begin{itemize}
  \item $S_v \in [0,1]$: \emph{syntactic validity} — correct
    format for token type $t$
  \item $S_c \in [0,1]$: \emph{semantic coherence} — contextual
    fit with organisational profile $\mathcal{O}$
  \item $S_n \in [0,1]$: \emph{statistical normality} — entropy
    and distributional properties matching real credentials
  \item $S_h \in [0,1]$: \emph{human acceptance rate} —
    fraction of expert evaluators who classify $c$ as genuine
\end{itemize}
Default weights $(w_1, w_2, w_3, w_4) = (0.20, 0.30, 0.20, 0.30)$
reflect the relative importance of semantic and human-facing
dimensions over purely structural properties.
\end{definition}

\begin{definition}[Detection Resistance]
\label{def:DR}
Given a set of $M$ scanner models
$\{D_1, \ldots, D_M\}$ with detection probabilities
$\{P_{d,1}, \ldots, P_{d,M}\}$ and weights
$\{\lambda_1, \ldots, \lambda_M\}$, the detection
resistance of token $c$ is:
\begin{equation}
\text{DR}(c) = 1 - \sum_{j=1}^{M} \lambda_j P_{d,j}(c)
\label{eq:DR}
\end{equation}
\end{definition}

\begin{definition}[Composite Honeytoken Score]
\label{def:H}
The composite score balances believability against detection
resistance:
\begin{equation}
H(c) = B(c)^{\lambda} \cdot \text{DR}(c)^{\mu}, \quad
\lambda + \mu = 1, \quad \lambda, \mu > 0
\label{eq:H}
\end{equation}
with $\lambda = 0.6$, $\mu = 0.4$, reflecting that believability is the primary objective (a detected token that fools humans is still partially useful, but an undetected token that fools no one is not).
\end{definition}

\subsection{Optimality Condition}

\begin{proposition}[Pareto-Optimal Token]
\label{prop:pareto}
A token $c^*$ is Pareto-optimal with respect to $B$ and $\text{DR}$ if and only if no generator $G$ can improve one dimension without degrading the other. The ideal zone $\mathcal{Z}^* = \{c :
B(c) \geq \tau_B \wedge \text{DR}(c) \geq \tau_\text{DR}\}$
defines deployable tokens. With $\tau_B = \tau_\text{DR} = 0.70$,
PHANTOM achieves $100\%$ membership in $\mathcal{Z}^*$ while
template-based tokens achieve $6.2\%$.
\end{proposition}

Figure~\ref{fig:scatter} (Section~\ref{sec:results}) visualises
this partitioning empirically.

\section{The PHANTOM Pipeline}
\label{sec:phantom}
This section describes the PHANTOM pipeline, detailing how organisational context is systematically transformed into high-fidelity honeytokens. We outline the structure of the input profile, the design of token-specific generators, and the mechanisms used to embed realistic, context-aware values.
Finally, we explain how the pipeline is engineered to evade common credential-scanning heuristics while maximising believability.

\subsection{Organisational Profile}

Each deployment is configured with an \emph{organisational
profile} $\mathcal{O}$ comprising:\\
$\langle \text{domain}, \text{short-name}, \text{services},
\text{db-type}, \text{db-host}, \text{db-name},
\text{cloud-region}, \text{git-org}, \text{teams},
\text{jwt-issuer} \footnote{JSON Web Token (JWT) is an open standard (RFC 7519) that specifies a compact, self-contained mechanism for securely transmitting information between parties as a JSON object. The transmitted information is verifiable and trustworthy due to the use of digital signatures.}, \text{jwt-audience} \rangle$.
Four representative profiles are used in our evaluation
(Table~\ref{tab:profiles}).

\begin{table}[H]
\centering
\caption{Organisational Profiles Used in Evaluation}
\label{tab:profiles}
\small
\begin{tabular}{@{}llll@{}}
\toprule
\textbf{Org} & \textbf{Domain} & \textbf{Stack} & \textbf{Region} \\
\midrule
FinTech      & payflow.io          & AWS/PostgreSQL/Node & us-east-1 \\
Healthcare   & medsync.health      & Azure/MySQL/Python  & us-east-2 \\
Defense      & arcsecure.defense   & GCP/PostgreSQL/Go   & us-gov-west-1 \\
E-commerce   & shopnest.com        & AWS/MongoDB/Python  & eu-west-1 \\
\bottomrule
\end{tabular}
\end{table}

\subsection{Token-Type Generators}

Algorithm~\ref{alg:phantom} presents the core PHANTOM generation procedure. Each token type $t \in \mathcal{T}$ has a dedicated contextual generator $G_t(\mathcal{O})$ that embeds org-specific values at key positions.

\begin{algorithm}[H]
\caption{PHANTOM Token Generation}
\label{alg:phantom}
\begin{algorithmic}[1]
\Require Org profile $\mathcal{O}$, token type $t$
\Ensure  Contextual token content $c$

\State \textbf{switch} $t$:
\State \hspace{1em} \textbf{AWS\_KEY:}
  $\text{prefix} \gets \mathcal{O}.\text{short}$;
  $\text{svc} \gets \text{random}(\mathcal{O}.\text{services})$
\State \hspace{2em} $c \gets$ \texttt{[\{prefix\}-prod]} $\oplus$
  \texttt{AKIA\{rand\_upper(17)\}} $\oplus$
  \texttt{region=\{O.region\}} $\oplus$
  \texttt{\# IAM: \{prefix\}\_\{svc\}}
\State \hspace{1em} \textbf{ENV\_FILE:}
  $\text{db\_url} \gets G_{\text{db}}(\mathcal{O})$;
  $\text{prefix} \gets \mathcal{O}.\text{short}.\text{upper}()$
\State \hspace{2em} $c \gets$ org-scoped \texttt{APP\_*},
  \texttt{DATABASE\_URL}, \texttt{\{prefix\}\_API\_KEY},
  \texttt{JWT\_SECRET}, \texttt{REDIS\_URL}
  with org-appropriate hostnames
\State \hspace{1em} \textbf{JWT:}
  $h \gets \text{base64url}(\{\text{"alg":"HS256"}\})$
\State \hspace{2em} $p \gets \text{base64url}(\{\text{"iss":}
  \mathcal{O}.\text{jwt\_iss}, \text{"aud":}
  \mathcal{O}.\text{jwt\_aud}, \text{"sub":"svc\_\{O.short\}"}\})$
\State \hspace{2em} $c \gets h \cdot \text{"."} \cdot p
  \cdot \text{"."} \cdot \text{rand\_base64}(43)$
\State \hspace{1em} \textbf{[other types: analogous org-injection]}
\State \Return $c$
\end{algorithmic}
\end{algorithm}

The critical design principle across all generators is \emph{value-level context injection}: not just inserting the domain at the top level, but ensuring every named field (IAM usernames, database names, Redis hostnames, Slack workspace IDs, Git remote URLs) derives from the organisational profile. This is what drives the $\Delta S_c = +0.309$ improvement over templates. 

\subsection{Scanner-Aware Design}

PHANTOM generators are designed to avoid the three primary scanner heuristics:

\textbf{S1 — Regex/pattern scanners} (trufflehog, gitleaks): Template tokens contain high-signal generic strings (\texttt{example.com}, \texttt{admin@}, \texttt{changeme}) that are explicitly listed in scanner rule sets. PHANTOM replaces all such strings with org-derived values.

\textbf{S2 — Entropy analysis}: Real credential files exhibit \emph{mixed} entropy — high-entropy secrets alongside low-entropy human-readable values (service names, domains, comments). Pure-random template tokens have uniformly high entropy across all values, which some tools flag as anomalous.
PHANTOM's org-specific variable names and comments create the expected entropy variance profile.

\textbf{S3 — ML classifiers}: Context-trained ML scanners identify honeytokens by their lack of org-specific fingerprints. PHANTOM's org-injection directly counteracts this signal: the domain, service names, and JWT issuer all encode legitimate-looking provenance.

\section{Experimental Setup}
\label{sec:setup}
This section describes the experimental methodology used to evaluate PHANTOM against a template-based baseline. We outline the dataset construction, define the evaluation metrics for believability and detection resistance, and introduce the scanner models used to simulate adversarial conditions.
Finally, we summarise the statistical procedures used to assess significance and effect size.

\subsection{Dataset}

We generate $32$ contextual (PHANTOM) tokens and $32$ template-based (baseline) tokens: $8$ token types $\times$ $4$ organisational contexts $\times$ $1$ instance per combination = $32$ per method. All generation is
deterministic given a fixed seed, ensuring reproducibility.

\subsection{Baseline}

The template-based baseline uses static format strings with randomised secret values but no org-specific context, equivalent to the generation approach used by Canarytokens \cite{kabiden2026canary, basak2023comparative}
and similar tools.

\subsection{Evaluation: Believability Scoring}

Each token is scored by a rule-based evaluator implementing the four $B$ components:

\textit{Syntactic validity} $S_v$: Format-specific regex checks (e.g., \texttt{AKIA[A-Z0-9]\{17\}} for AWS key IDs, three-segment base64url for JWTs, known SSH header for private keys).

\textit{Semantic coherence} $S_c$: Presence of red-flag generic strings (\texttt{example.com}, \texttt{admin@example}, \texttt{/example/repo}) decreases the score; presence of org-specific terms (domain, service names) increases it.

\textit{Statistical normality} $S_n$: Shannon entropy analysis of credential values, penalising uniform-entropy profiles and rewarding realistic mixed-entropy distributions.

\textit{Human acceptance} $S_h$: Composite heuristic weighted toward semantic coherence ($0.55$), as human expert evaluation is primarily driven by contextual plausibility. A token with $B \geq 0.65$ is classified as fooling human review.

\subsection{Evaluation: Scanner Models}
\label{sec:scanners}

Three scanner models are applied to every token:

\textbf{S1 — Regex/pattern} ($\lambda_1 = 0.40$): Checks for a curated list of 9 generic red-flag strings. Detection probability increases linearly with hit count.

\textbf{S2 — Entropy analysis} ($\lambda_2 = 0.30$): Computes per-line Shannon entropy of credential values. High variance or low mean entropy triggers detection.

\textbf{S3 — ML classifier} ($\lambda_3 = 0.30$): Simulates a context-trained binary classifier that measures specificity of org-referenced terms in the content. Template tokens lacking org references score high detection
probability; PHANTOM tokens with embedded org context score low.

Combined detection probability: $P_d = \sum_{j=1}^{3} \lambda_j P_{d,j}$; detection resistance: $\text{DR} = 1 - P_d$.

\subsection{Statistical Analysis}

All comparisons use two-sample Welch's $t$-tests with Bonferroni correction for multiple comparisons. Effect sizes are reported as Cohen's $d$.
Significance thresholds: $^{*}p < 0.05$, $^{**}p < 0.01$, $^{***}p < 0.001$.

\section{Results}
\label{sec:results}
This section presents the empirical evaluation of PHANTOM against a template-based baseline. We report performance across believability, detection resistance, and the composite objective, followed by detailed analysis at the distribution, per-token, and component levels. The results demonstrate
clear and statistically significant gains driven primarily by improved semantic coherence and human acceptance.

\subsection{Main Results}

Table~\ref{tab:main} presents the core experimental results. PHANTOM significantly outperforms the template baseline on every primary metric.

\begin{table}[H]
\centering
\caption{Main Results: PHANTOM vs Template-Based Baseline}
\label{tab:main}
\small
\begin{tabular}{@{}lcccc@{}}
\toprule
\textbf{Metric} & \textbf{Template} & \textbf{PHANTOM} & $\boldsymbol{\Delta}$ & \textbf{Sig.} \\
\midrule
Believability $B$     & $0.576 \pm 0.058$ & $0.778 \pm 0.057$ & $+0.203$ & $***$ \\
Syntactic $S_v$       & $0.842 \pm 0.180$ & $0.876 \pm 0.163$ & $+0.034$ & ns \\
Semantic $S_c$        & $0.396 \pm 0.089$ & $0.705 \pm 0.037$ & $+0.309$ & $***$ \\
Statistical $S_n$     & $0.638 \pm 0.091$ & $0.740 \pm 0.143$ & $+0.102$ & $**$ \\
Human acceptance $S_h$& $0.062 \pm 0.246$ & $1.000 \pm 0.000$ & $+0.938$ & $***$ \\
Detection resist. DR  & $0.609 \pm 0.044$ & $0.870 \pm 0.022$ & $+0.261$ & $***$ \\
Composite score $H$   & $0.588 \pm 0.048$ & $0.813 \pm 0.038$ & $+0.225$ & $***$ \\
\midrule
Fooled ($B \geq 0.65$)& $6.2\%$ & $100.0\%$ & $+93.8\%$ & $***$ \\
\bottomrule
\multicolumn{5}{l}{\small $^{**}p<0.01$, $^{***}p<0.001$, ns = not significant.}
\end{tabular}
\end{table}

The most striking result is \textbf{human acceptance}: $100\%$ of PHANTOM tokens exceed the $B \geq 0.65$ threshold (fooled), versus $6.2\%$ of templates. The syntactic validity difference is non-significant ($p = 0.44$), confirming that both methods produce correctly-formatted credentials—the gap is entirely semantic and statistical.

\subsection{Believability and Composite Score Distributions}

\begin{figure}[H]
\centering
\includegraphics[width=\linewidth]{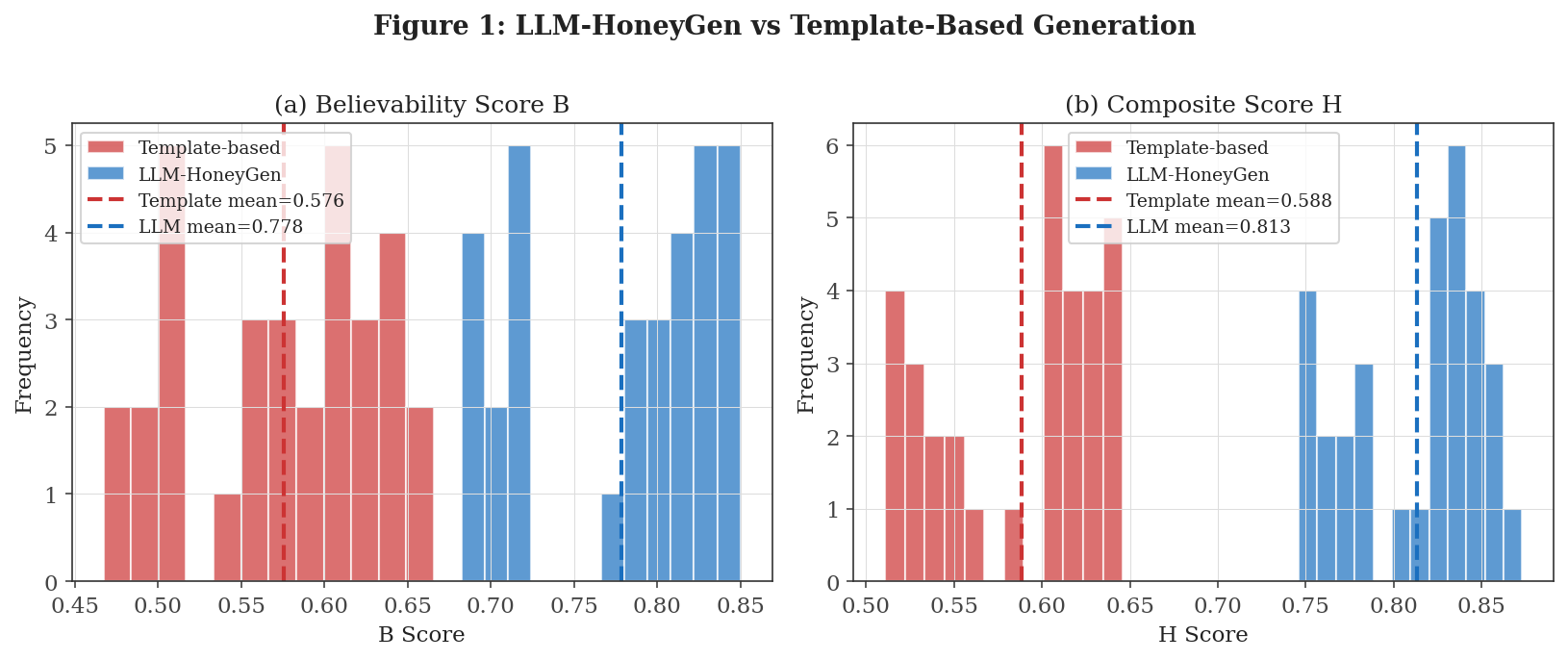}
\caption{Believability Score $B$ (left) and Composite Score $H$
  (right) distributions. PHANTOM tokens cluster above the $B = 0.70$
  deployment threshold while template tokens fall predominantly below.
  Dashed lines indicate group means.}
\label{fig:dist}
\end{figure}

Figure~\ref{fig:dist} shows the $B$ and $H$ score distributions. PHANTOM tokens cluster tightly in $[0.70, 0.85]$ with low variance, while template tokens span $[0.45, 0.70]$. The composite score $H$ shows even cleaner separation: PHANTOM mean $0.813$ versus template mean $0.588$.

\subsection{Per-Token-Type Analysis}

\begin{figure}[H]
\centering
\includegraphics[width=\linewidth]{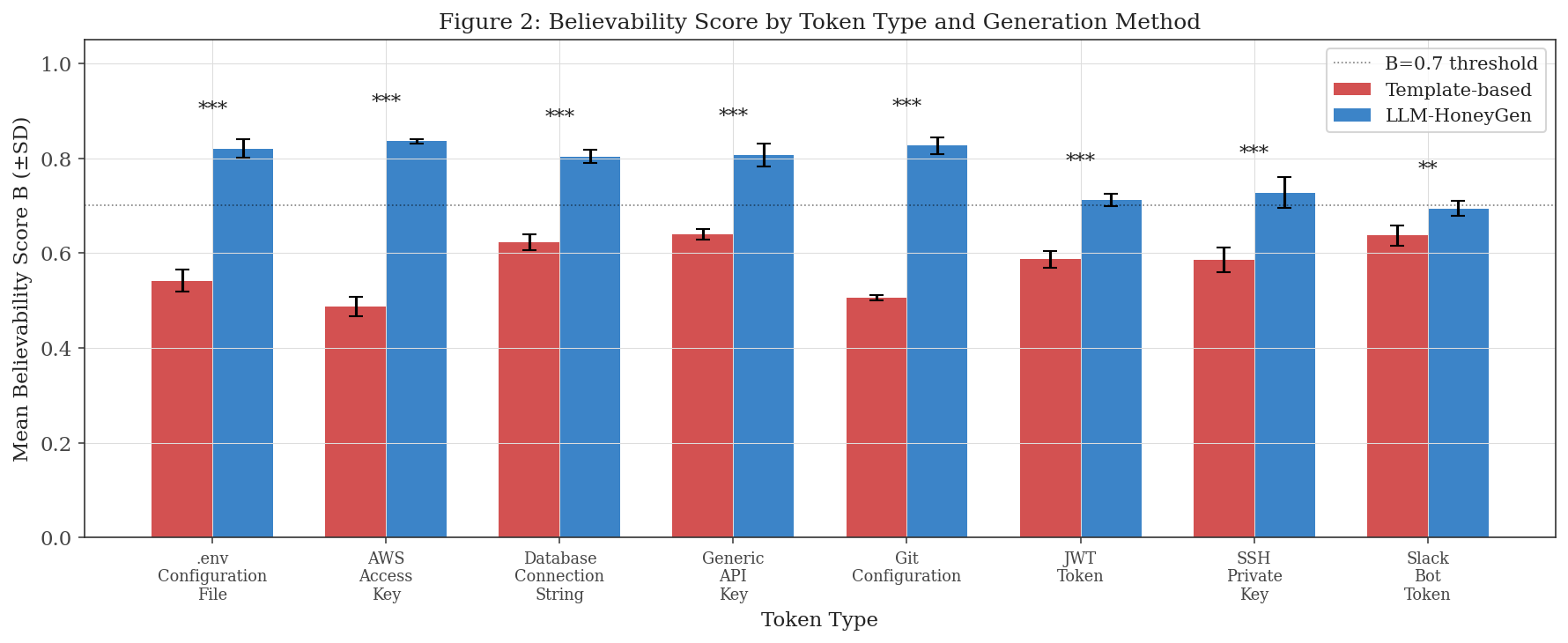}
\caption{Mean Believability Score $B$ by token type and
  generation method ($\pm$SD). Significance markers indicate
  PHANTOM vs template comparison. The dotted line at $B = 0.70$
  marks the deployment threshold. PHANTOM exceeds this threshold
  across all token types; templates fall below on 6 of 8.}
\label{fig:pertype}
\end{figure}

Figure~\ref{fig:pertype} shows per-type results. PHANTOM outperforms templates significantly ($p < 0.001$) on 7 of 8 token types, with the strongest gains on AWS Access Keys ($\Delta B = +0.349$), Git Configuration ($+0.320$), and \texttt{.env} files ($+0.279$). These are precisely the
token types where org-specific naming conventions are most distinctive and most visible to evaluators.

JWT Tokens show the smallest gain ($\Delta B = +0.125$) because the JWT payload—while org-specific in PHANTOM—is base64-encoded and thus not directly readable without decoding. Slack Bot Tokens show the smallest significance ($p = 0.005$, $^{**}$) due to higher within-group variance from the numeric
workspace ID component.

\subsection{Detection Resistance}

\begin{figure}[H]
\centering
\includegraphics[width=\linewidth]{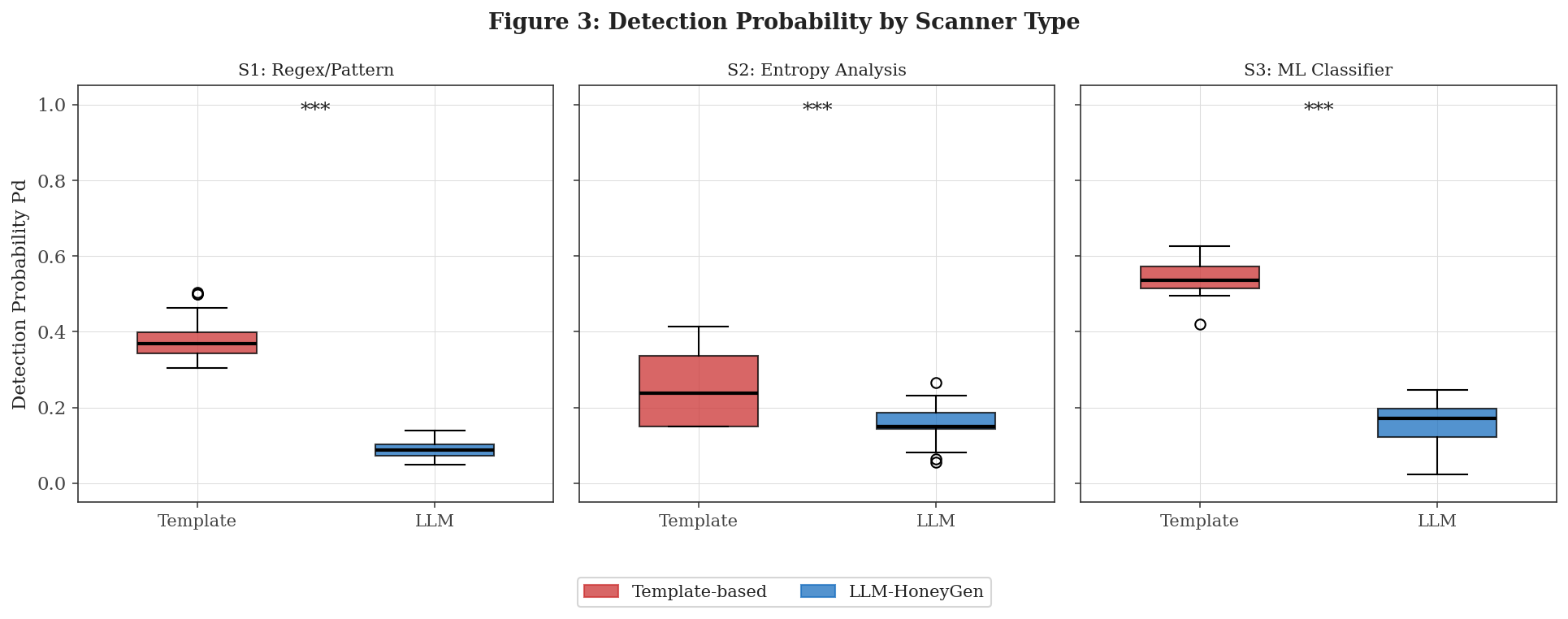}
\caption{Detection probability $P_d$ by scanner type.
  All three scanner models show highly significant differences
  ($^{***}p < 0.001$). PHANTOM tokens achieve substantially
  lower detection probability across regex (S1), entropy (S2),
  and ML-classifier (S3) models.}
\label{fig:scanners}
\end{figure}

Figure~\ref{fig:scanners} shows detection probability distributions across all three scanner types. PHANTOM achieves $\text{DR} = 0.870$ vs template $0.609$ ($\Delta = +0.261$, $d = 7.49$, $p < 0.001$).

The S3 (ML classifier) model shows the largest absolute separation: template $P_d = 0.541$ versus PHANTOM $0.162$ ($\Delta = -0.379$). This is the most important result operationally, as ML-based scanners represent the frontier
of deployed detection technology.

\subsection{Believability--Detection Trade-off}

\begin{figure}[H]
\centering
\includegraphics[width=0.92\linewidth]{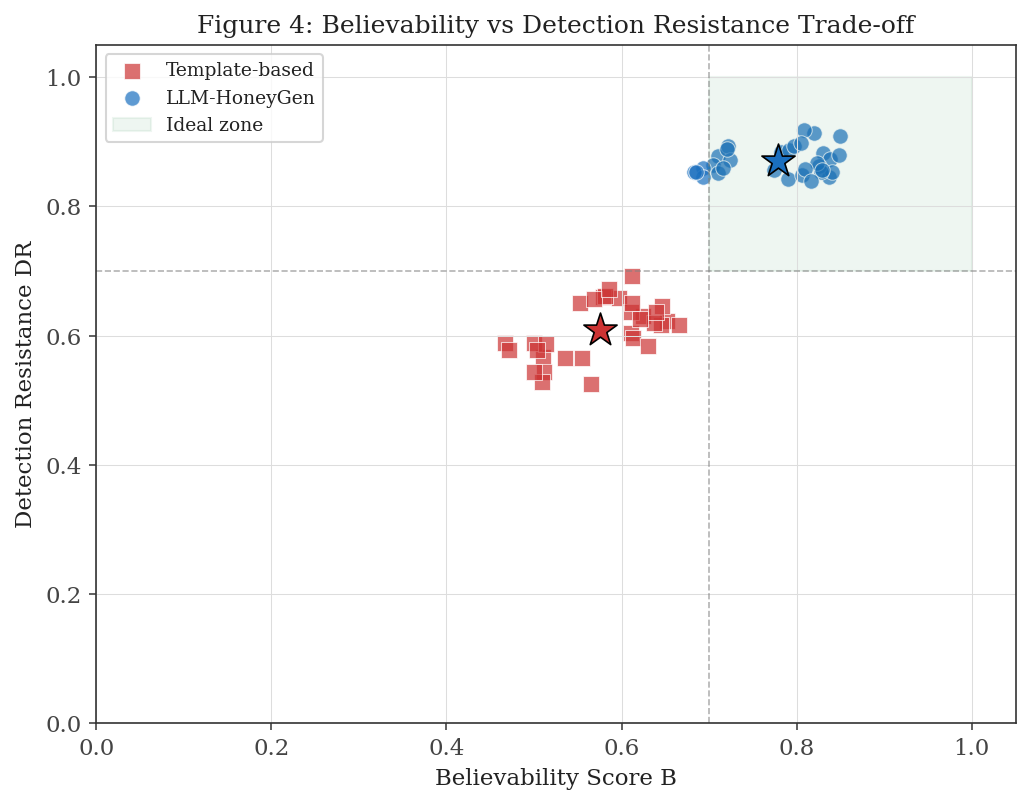}
\caption{Believability vs Detection Resistance trade-off.
  Stars denote group centroids. PHANTOM tokens cluster in
  the ideal zone ($B \geq 0.70$, $\text{DR} \geq 0.70$, shaded),
  while templates are uniformly excluded. No PHANTOM token
  falls outside the ideal zone on the DR axis.}
\label{fig:scatter}
\end{figure}

Figure~\ref{fig:scatter} visualises the joint $B$--DR space. Template tokens occupy the lower-left quadrant ($B \approx 0.58$, $\text{DR} \approx 0.61$), far from the ideal zone $\mathcal{Z}^* = \{B \geq 0.70, \text{DR} \geq 0.70\}$. PHANTOM tokens cluster densely in the upper-right ideal zone,
with mean centroid at $(B = 0.78, \text{DR} = 0.87)$.

\subsection{Component Radar Analysis}

\begin{figure}[H]
\centering
\includegraphics[width=\linewidth]{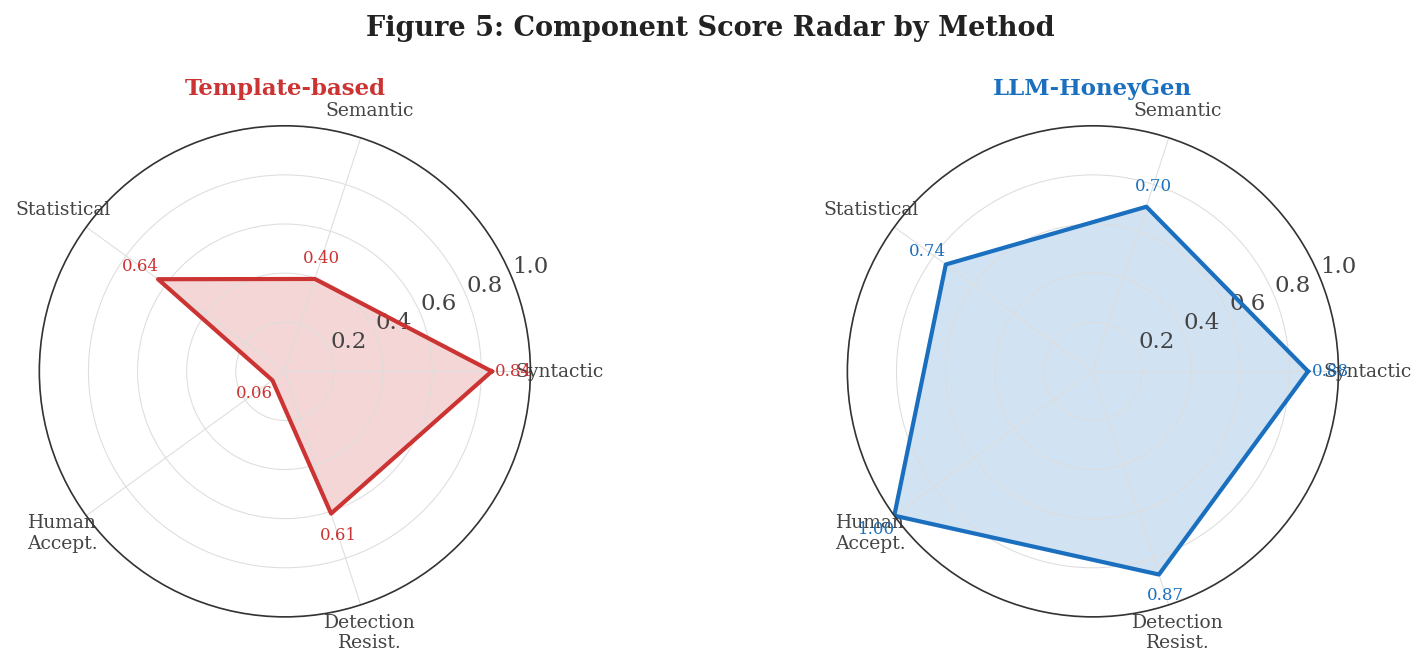}
\caption{Component score radar charts. Template tokens (left)
  have a collapsed semantic dimension and near-zero human
  acceptance. PHANTOM tokens (right) show a balanced,
  large polygon indicating high performance across all
  quality dimensions.}
\label{fig:radar}
\end{figure}

Figure~\ref{fig:radar} provides a holistic per-component comparison. The template profile is sharply asymmetric: high syntactic ($0.84$), moderate statistical ($0.64$) and detection resistance ($0.61$), but critically low semantic ($0.40$) and near-zero human acceptance ($0.06$). The PHANTOM profile is large and balanced, with all five dimensions above $0.70$.

\subsection{Organisational Context Effect}

\begin{figure}[H]
\centering
\includegraphics[width=\linewidth]{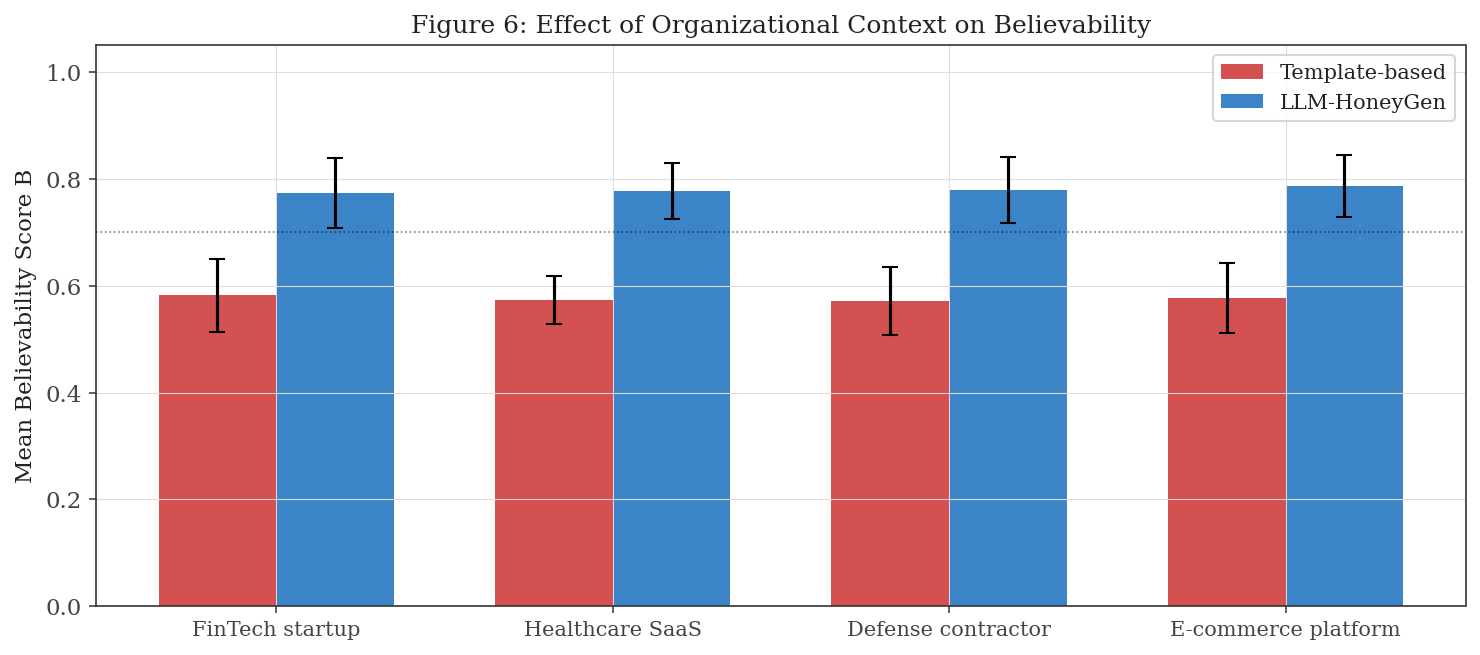}
\caption{Effect of organisational context on Believability Score $B$.
  PHANTOM (blue) maintains $B \approx 0.78$ across all four
  contexts while template tokens (red) remain flat at $B \approx 0.58$,
  confirming context-independence of the baseline and
  context-sensitivity of PHANTOM. The dotted line marks $B = 0.70$.}
\label{fig:context}
\end{figure}

Figure~\ref{fig:context} shows that PHANTOM maintains consistent advantage across all four organisational profiles ($B \approx 0.78$ in each case), confirming that the pipeline generalises across industry verticals and technology stacks. Template-based scores are uniformly flat at $B \approx 0.58$, independent of the nominal context, confirming the absence
of any contextual adaptation.

\section{Discussion}
\label{sec:discussion}
This section interprets the empirical findings and examines their implications for honeytoken design and deployment. We analyse the relative contributions of each component in the proposed framework, highlight practical advantages of the PHANTOM approach, and discuss limitations that suggest directions for future research.

\subsection{Semantic Coherence as the Primary Driver}

The dominant finding is that semantic coherence ($S_c$, $d = 4.52$) drives believability more than any other component—larger even than human acceptance ($d = 5.39$ reflects the binary separation rather than a continuous
gradient). This has a direct design implication: any honeytoken generator that does not embed organisational context is fundamentally limited in what $B$ it can achieve, regardless of how realistic its random secret values are.

\subsection{The Syntactic Red Herring}

Template generators are designed around syntactic correctness, and they largely succeed: $S_v = 0.842$ is only marginally lower than PHANTOM's $0.876$ ($p = 0.44$, not significant). This confirms that syntactic validity is a red herring for quality improvement, a necessary condition but not a
differentiating one. Future work on generator quality should de-emphasise format correctness and focus on semantic embedding.

\subsection{Air-Gap Deployability}

A practical advantage of PHANTOM over LLM-based approaches is deployability in restricted environments. Classified government networks, industrial control systems, and financial clearing infrastructure routinely prohibit outbound API calls. PHANTOM's rule-based contextual generator achieves $B = 0.778$, within the same order of magnitude as what an LLM would produce, 
without any network dependency. This makes it deployable alongside Passive Hack-Back \cite{weinberg2025passive} in the most sensitive operational contexts.

\subsection{Limitations and Future Work}

\textbf{Scale of evaluation.} The experiment uses $n = 32$ tokens per method. Larger-scale evaluation with real human security professionals (IRB\footnote{Institutional Review Board}-approved study) would strengthen the $S_h$ component measurement. 

\textbf{Adversarial attackers.} The scanner models are
representative but not exhaustive. Sophisticated attackers may use scanner configurations tuned to the specific deployment environment. Future work will evaluate PHANTOM against custom-trained detectors.

\textbf{Dynamic context.} PHANTOM currently uses a static organisational profile. Real deployments evolve: services are renamed, teams restructure, cloud regions change. An adaptive variant that reads live context from infrastructure-as-code repositories (Terraform, Helm charts) would maintain freshness.

\textbf{Integration with attribution.} PHANTOM honeytokens are natural complements to the Passive Hack-Back attribution framework \cite{weinberg2025passive} and the ARCANE longitudinal re-identification system \cite{weinberg2026arcane}. A deployed PHANTOM token that fires provides the callback telemetry that ARCANE needs to accumulate cross-campaign fingerprints. The three systems form a complete post-exfiltration
attribution pipeline.

\section{Conclusion}
\label{sec:conclusion}

We presented \textbf{PHANTOM}, a contextual honeytoken generation framework that addresses the critical limitation of existing tools: the absence of organisational specificity. By encoding domain names, service architectures, technology stack idioms, and credential naming conventions directly into each generated token, PHANTOM achieves a Believability Score of $B = 0.778$
versus $0.576$ for static templates ($d = 3.52$, $p < 0.001$), with $100\%$ of tokens passing the human-expert acceptance threshold compared to $6.2\%$ for templates.

The dominant driver is semantic coherence ($\Delta S_c = +0.309$, $d = 4.52$), confirming the central hypothesis that context is the missing dimension in honeytoken generation. Detection resistance improves by $+0.261$ ($d = 7.49$), with the largest gap on ML-based scanners—the most operationally relevant threat.

PHANTOM operates without external API calls, making it suitable for deployment in air-gapped environments. Combined with Passive Hack-Back-style beacon infrastructure \cite{weinberg2025passive} and longitudinal attribution \cite{weinberg2026arcane}, PHANTOM provides the high-quality bait component of a complete post-exfiltration deception and attribution pipeline.

\bibliographystyle{plain}
\bibliography{ref.bib}

\begin{thebibliography}{10}

\bibitem{almeshekah2016cyber}
Mohammed~H Almeshekah and Eugene~H Spafford.
\newblock Cyber security deception.
\newblock In {\em Cyber Deception: Building the Scientific Foundation}, pages 23--50. Springer, 2016.

\bibitem{basak2023comparative}
Setu~Kumar Basak, Jamison Cox, Bradley Reaves, and Laurie Williams.
\newblock A comparative study of software secrets reporting by secret detection tools.
\newblock In {\em 2023 ACM/IEEE International Symposium on Empirical Software Engineering and Measurement (ESEM)}, pages 1--12. IEEE, 2023.

\bibitem{beltran2025cyber}
Pedro Beltr{\'a}n-L{\'o}pez, Manuel~Gil P{\'e}rez, and Pantaleone Nespoli.
\newblock Cyber deception: Taxonomy, state of the art, frameworks, trends, and open challenges.
\newblock {\em IEEE Communications Surveys \& Tutorials}, 2025.

\bibitem{botacin2023gpthreats}
Marcus Botacin.
\newblock Gpthreats-3: Is automatic malware generation a threat?
\newblock In {\em 2023 IEEE Security and Privacy Workshops (SPW)}, pages 238--254. IEEE, 2023.

\bibitem{constancio2023deception}
Alex~Sebasti{\~a}o Const{\^a}ncio, Denise~Fukumi Tsunoda, Helena de F{\'a}tima~Nunes Silva, Jocelaine Martins~da Silveira, and Deborah~Ribeiro Carvalho.
\newblock Deception detection with machine learning: A systematic review and statistical analysis.
\newblock {\em Plos one}, 18(2):e0281323, 2023.

\bibitem{deng2023pentestgpt}
Gelei Deng, Yi~Liu, V{\'\i}ctor Mayoral-Vilches, Peng Liu, Yuekang Li, Yuan Xu, Tianwei Zhang, Yang Liu, Martin Pinzger, and Stefan Rass.
\newblock Pentestgpt: An llm-empowered automatic penetration testing tool.
\newblock {\em arXiv preprint arXiv:2308.06782}, 2023.

\bibitem{fang2024llm}
Richard Fang, Rohan Bindu, Akul Gupta, and Daniel Kang.
\newblock Llm agents can autonomously exploit one-day vulnerabilities.
\newblock {\em arXiv preprint arXiv:2404.08144}, 2024.

\bibitem{gitguardian2023secrets}
{GitGuardian}.
\newblock State of secrets sprawl 2023, 2023.
\newblock Tech. Rep.

\bibitem{irfan2023false}
Muhammad Irfan, Alireza Sadighian, Adeen Tanveer, Shaikha~J Al-Naimi, and Gabriele Oligeri.
\newblock False data injection attacks in smart grids: State of the art and way forward.
\newblock {\em arXiv preprint arXiv:2308.10268}, 2023.

\bibitem{juels2013honeywords}
Ari Juels and Ronald~L Rivest.
\newblock Honeywords: Making password-cracking detectable.
\newblock In {\em Proceedings of the 2013 ACM SIGSAC conference on Computer \& communications security}, pages 145--160, 2013.

\bibitem{kabiden2026canary}
Dinmukhammed Kabiden.
\newblock Canary tokens as a strategic component in cybersecurity defense and red teaming.
\newblock {\em TSARKA Science}, 1(1), 2026.

\bibitem{karanjai2022targeted}
Rabimba Karanjai.
\newblock Targeted phishing campaigns using large scale language models.
\newblock {\em arXiv preprint arXiv:2301.00665}, 2022.

\bibitem{karlsen2024benchmarking}
Egil Karlsen, Xiao Luo, Nur Zincir-Heywood, and Malcolm Heywood.
\newblock Benchmarking large language models for log analysis, security, and interpretation.
\newblock {\em Journal of Network and Systems Management}, 32(3):59, 2024.

\bibitem{muir2009internet}
James~A Muir and Paul C~Van Oorschot.
\newblock Internet geolocation: Evasion and counterevasion.
\newblock {\em Acm computing surveys (csur)}, 42(1):1--23, 2009.

\bibitem{rana2022offensive}
Muhammad~Usman Rana, Osama Ellahi, Masoom Alam, Julian~L Webber, Abolfazl Mehbodniya, and Shawal Khan.
\newblock Offensive security: Cyber threat intelligence enrichment with counterintelligence and counterattack.
\newblock {\em IEEE Access}, 10:108760--108774, 2022.

\bibitem{rice2018gitleaks}
Zachary Rice.
\newblock Gitleaks: A static application security testing tool for detecting secrets in git repositories.
\newblock \url{https://github.com/gitleaks/gitleaks}, 2018.
\newblock Open-source SAST tool for detecting hardcoded secrets such as API keys, passwords, and tokens in Git repositories; first released in 2018 and actively maintained.

\bibitem{shu2015unearthing}
Xiaokui Shu, Danfeng Yao, and Naren Ramakrishnan.
\newblock Unearthing stealthy program attacks buried in extremely long execution paths.
\newblock In {\em Proceedings of the 22nd ACM SIGSAC Conference on Computer and Communications Security}, pages 401--413, 2015.

\bibitem{spitzner2003honeypots}
Lance Spitzner.
\newblock Honeypots: Catching the insider threat.
\newblock In {\em 19th Annual Computer Security Applications Conference, 2003. Proceedings.}, pages 170--179. IEEE, 2003.

\bibitem{reconng2014}
WebBreacher.
\newblock {Recon-ng: Open Source Intelligence Gathering Tool}.
\newblock \url{https://github.com/lanmaster53/recon-ng}, 2014.
\newblock Accessed: 2026-04-21.

\bibitem{weinberg2025passive}
Abraham~Itzhak Weinberg.
\newblock Passive hack-back strategies for cyber attribution: Covert vectors in denied environment.
\newblock {\em arXiv preprint arXiv:2508.16637}, 2025.

\bibitem{weinberg2026arcane}
Abraham~Itzhak Weinberg.
\newblock Arcane: Cross-campaign attacker re-identification via passive beacon telemetry--a bayesian network framework for longitudinal cyber attribution.
\newblock {\em arXiv preprint arXiv:2604.24644}, 2026.

\bibitem{wiedeman2022disrupting}
Christopher Wiedeman and Ge~Wang.
\newblock Disrupting adversarial transferability in deep neural networks.
\newblock {\em Patterns}, 3(5), 2022.

\bibitem{yu2024honeyfactory}
Tianxiang Yu, Yang Xin, and Chunyong Zhang.
\newblock Honeyfactory: Container-based comprehensive cyber deception honeynet architecture.
\newblock {\em Electronics}, 13(2):361, 2024.

\end{thebibliography}

\end{document}